\documentclass{llncs}
\usepackage{url}
\usepackage{subfig}
\usepackage{graphicx}
\usepackage{epstopdf} % Allow EPS figures to be included in PDF output
\DeclareGraphicsExtensions{.pdf} % {.png,.pdf,.ps,.eps}
\graphicspath{{figs/}}

\title{Synaptic Scaling Balances Learning in a Spiking Model of Neocortex}
\titlerunning{Scaling Balances Learning in Neocortex}

\author{Mark Rowan\inst{1} \and Samuel Neymotin\inst{2}}

\institute{School of Computer Science, University of Birmingham, UK
\email{m.s.rowan@cs.bham.ac.uk} \and Dept. Neurobiology, Yale University School
of Medicine, New Haven, USA \email{samuel.neymotin@yale.edu}}

\begin{document}
\maketitle
\begin{abstract}
Learning in the brain requires complementary mechanisms: potentiation and
activity-dependent homeostatic scaling. We introduce synaptic
scaling to a biologically-realistic spiking model of neocortex which can learn
changes in oscillatory rhythms using STDP, and show that scaling is necessary
to balance both positive and negative changes in input from potentiation and
atrophy. We discuss some of the issues that arise when considering synaptic
scaling in such a model, and show that scaling regulates activity whilst
allowing learning to remain unaltered.
\end{abstract}

% Introduction
\section{Introduction}
Spike Timing-Dependent Plasticity (STDP), a phenomenological learning rule in
which synaptic potentiation and depression depend upon relative firing times
\cite{dan2004spike,zhang1998critical}, has been used to learn oscillatory
rhythms in neocortical models. In an existing biologically-realistic spiking
model of neocortex \cite{neymotin2011emergence}, applying excitatory to
excitatory (E$\to$E) STDP with a rhythmic training signal led to
hyper-potentiation through positive feedback: strengthened synapses drove
postsynaptic neurons to fire immediately, leading to further potentiation. This
unbounded potentiation then pushed the network into synchronized epileptiform
firing. Directly opposing E$\to$E learning with equal excitatory to inhibitory
(E$\to$I) potentiation partially balanced this positive feedback. However,
epileptiform behaviour still occurred with high-frequency signals
\cite{neymotin2011training}.

We postulated that a homeostatic mechanism might be a solution to this problem.
Neuronal homeostatic synaptic scaling is a local feedback mechanism which senses
levels of activity-dependent cytosolic calcium within the cell and adjusts
neuronal firing activity accordingly. This is achieved by producing alterations
in excitatory AMPA receptor accumulation in response to changes in firing
activity occurring over hours to days \cite{turrigiano2008self}, leading to
changes in the excitability of the neuron.

During learning, synaptic scaling plays an important role in balancing
potentiation. By constantly shifting mean activation back towards a target
activity level, but maintaining the learned relative distribution of presynaptic
weights, global levels of activity can be regulated \cite{vanrossum2000stable}.
During periods of hypoactivity (e.g. in degenerative disorders), synaptic
scaling is also capable of raising the sensitivity of neurons via AMPA receptor
upregulation, so that activity levels can be restored \cite{turrigiano2008self}.

Previous work has demonstrated synaptic scaling with learning in a single-neuron
model \cite{vanrossum2000stable}. It has also been shown that synaptic scaling
can prevent input saturation in a spiking neural network in the absence of
learning \cite{chandler2012joining}. In this paper, we add long-term synaptic
plasticity to a spiking neural network to show that homeostatic synaptic scaling
can prevent hyper-potentiation while preserving learned information.

\section{Methods}
The model was based on the anatomy of a single column of sensory neocortex
\cite{neymotin2011emergence,binzegger2004quantitative,lefort2009excitatory}. It
was composed of 470 neurons divided into 3 types (excitatory pyramidal cells E,
fast-spiking interneurons I, and low-threshold spiking interneurons IL),
distributed across the 6 layers of the neocortex. This yielded 13 neuronal
populations in total, with the following numbers of cells per type: E2 (i.e.
excitatory layer 2/3 cell), 150; I2 (fast spiking interneuron in layer 2/3), 25;
I2L (low-threshold spiking interneuron in layer 2/3), 13; E4, 30; I4, 20; I4L,
14; E5a, 65; E5b, 17; I5, 25; I5L, 13; E6, 60; I6, 25; and I6L, 13.

The cell model was an extension of an integrate-and-fire unit with added
complexity (adaptation, bursting, depolarization blockade, and voltage-sensitive
NMDA conductance) in the form of rules  \cite{lytton2006rule}, and was simulated
in an event-driven fashion where cell state variables were only calculated at
input events, making use of previously developed just-in-time synapses optimized
for large networks supporting high-frequency synaptic events
\cite{lytton2008just}. Each cell had fast inhibitory GABA$_A$ receptors, fast
excitatory AMPA receptors, and slow excitatory NMDA receptors, with each
producing a voltage-step with following decay.

In addition to spikes generated by cells in the model, subthreshold
Poisson-distributed spike inputs to each synapse were used to maintain activity
in the model: 100--150~Hz for GABA$_A$, 240--360~Hz for AMPA receptors, and
40--60~Hz for NMDA receptors. These external inputs represented the inputs from
other regions of the brain. To simulate additional afferent sensory inputs,
low-amplitude training signals were applied to the layer 4 excitatory neurons
(E4) in some simulations. STDP was implemented on AMPA synapses from E$\to$E
cells using a basic model with bilateral exponential decay (40ms maximal
interspike difference, 10ms time constant) incrementing by 0.1\% of baseline
synaptic weight. It should be noted that STDP in this model is additive, whilst
van Rossum argues that it should be multiplicative \cite{vanrossum2000stable}.
Further details of the cell model can be found in \cite{neymotin2011emergence}
and \cite{neymotin2011training}.
% %%%% SN - is the value 0.1% of baseline or 1% of baseline? if plastEEinc ==
% 0.01 that means the increment is 1% of baseline weight.

% %%% MR - no, plastEEinc == 0.001, i.e. 0.1% increment. Any larger and it
% pushes the network to epilepsy (albeit only briefly with scaling enabled, as
% scaling soon recovers normal activity) but even with E->I STDP activated I
% couldn't replicate the results in the STDP paper, which I think may be down to
% the bug where some weights were not getting summed properly, which you told me
% was fixed in INTF6 after publication.

Scaling was implemented at E cell AMPA synapses by multiplying each cell $i$'s
postsynaptic input by a scale factor $w_i$, representing the multiplicative
accumulation of AMPA receptors at synapses. Changes in the scale factor were
calculated following the formula of van Rossum et al.
\cite{vanrossum2000stable}, with $a_i$ as the cell's firing activity,
$a_i^{goal}$ as the target activity, $\beta$ as the scaling strength, $\gamma$
as the ``integral controller'' weight, and $\frac{dw_i(t)}{dt}$ as the rate of
change of the synaptic weight:
\begin{equation}
\frac{dw_i(t)}{dt} = \beta w_i(t) [a_i^{goal} - a_i(t)]
+ \gamma w_i(t) \int_0^t{dt'[a_i^{goal} - a_i(t')]}
\label{eq:scaling}
\end{equation}
The following parameter values were used: strength
$\beta=4.0\times10^{-8}$/ms/Hz; integral controller weight
$\gamma=1.0\times10^{-10}$/ms$^2$/Hz; activity sensor time constant
$\tau=100\times10^3$ ms. Scaling was applied inversely at GABA$_A$ synapses
(i.e. by multiplying postsynaptic input by $\frac{1}{w_i}$) to enable the
scaling of excitatory and inhibitory synapses in opposite directions, mimicking
the effect of global growth factors such as BDNF
\cite{turrigiano2008self,chandler2012joining,rutherford1998bdnf,turrigiano2011too}.

%%%% WL -- this weakens a little the dichotomy we drew in the intro between balancing
%%%%       E>E with E>I (or I>E) and homeostasis, but really doesn't matter
%%%%% SN - so you do have plasticity at gabaa synapses?? I thought you had
%%%%% mentioned you didn't have plasticity there but I must be mistaken
%%%% MR - Yes, see intf6.mod:592. GABAA synapses on E-cells are scaled in the
%%%% opposite direction to AMPA synapses, which is experimentally supported.
%%%% There's no scaling for I-cells (by default; this can be turned on using
%%%% setIscaling() if desired, but as mentioned in the Discussion, it destabilises
%%%% the network).

Average activity level for each cell $i$ was sensed using van Rossum's
slow-varying sensor $a_i(t)$, which increased monotonically with spike $t_x$ at
current timestep $t$, and decayed otherwise \cite{vanrossum2000stable}:
\begin{equation}
\tau \frac{da_i(t)}{dt} = -a_i(t) + \sum_x{\delta(t-t_x)}
\label{eq:constsensor}
\end{equation}
The sensor decays exponentially as it is updated at each non-firing timestep.
However, the use of event-driven just-in-time synapses
\cite{neymotin2011training,lytton2008just} meant that cell states were only
updated upon each spike event rather than at every timestep, so inter-spike
decay of the activity sensor could only be calculated periodically. We therefore
modified the activity sensor. Here, the first term decays the sensor according
to the time between spikes $t-t_x$, and the second term increments it for the
new spike, with both terms updated concurrently on the occurrence of a spike
at time $t_x$:
\begin{equation}
a_i(t) = a_i(t_x) e^{-\frac{1}{\tau}(t - t_x)} + \frac{1 - a_i(t_x)}{\tau}
\label{eq:periodicsensor}
\end{equation}

\begin{figure}[t]
\begin{center}
  \includegraphics[width=\textwidth]{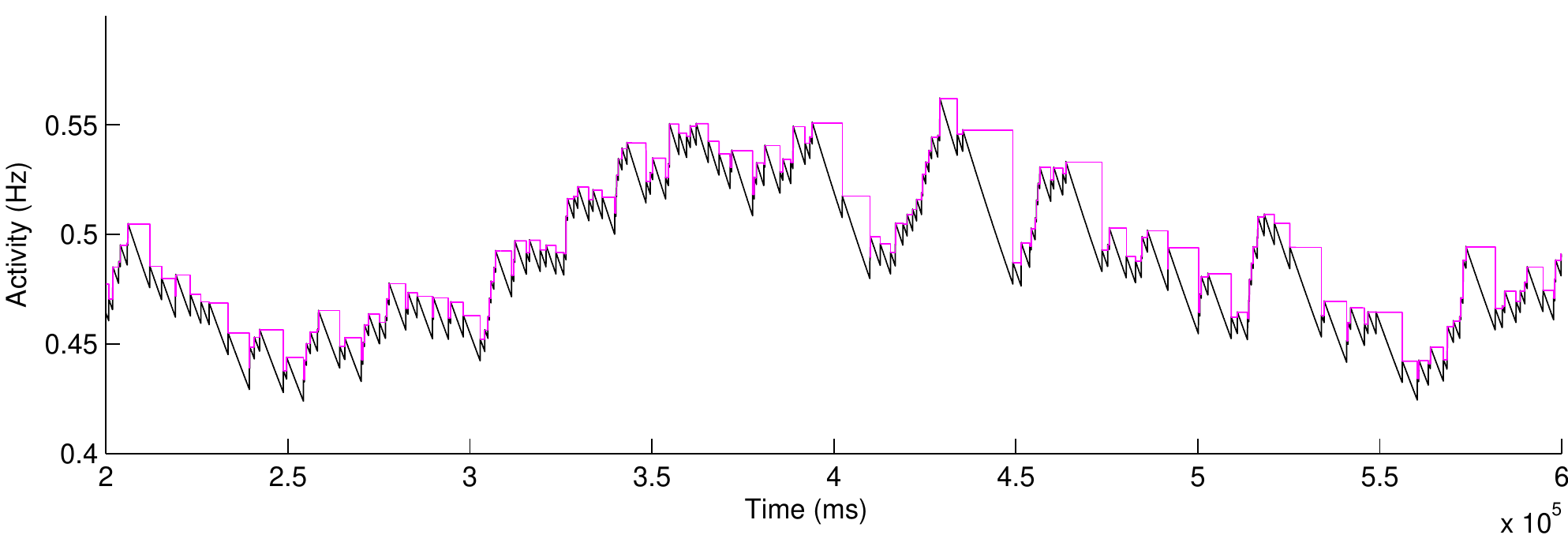}
  \caption{Activity sensor updating at every simulation timestep
  (Eqn.~\ref{eq:constsensor}; black) and at every spike for activity-driven
  just-in-time synapses (Eqn.~\ref{eq:periodicsensor}; magenta).}
  \label{fig:actsensor}
\end{center}
\end{figure}

Figure \ref{fig:actsensor} shows the activity of a simulated, randomly-spiking
neuron operating under the constant-timestep update policy
(\ref{eq:constsensor}), and the equivalent activity values under the
periodic-update policy (\ref{eq:periodicsensor}). The activity rises identically
in both cases when spikes occur, but the periodic sensor does not decay until
the next spike event occurs, giving the step-like appearance. The values at the
spike times are correct down to round-off error at the spike times.

Instead of providing an arbitrary rate target for each cell, which would
fundamentally affect network dynamics, the intrinsic dynamics of the network
were used to provide set-points. Initially, with synaptic scaling off, activity
sensors began at 0~Hz.  They were then adjusted over 800~s of simulated time
based on the activity level of the cells. Synaptic scaling was then switched on.

A time constant $\tau$ of 100~s \cite{vanrossum2000stable} leads to a
simulation timescale of several hours for synaptic scaling: far closer to the
expected biological timescale than previous studies
\cite{turrigiano2008self,frohlich2008pathological,chandler2012joining}. To
achieve this length of simulation, the model was extended to allow periodic
flushing of all spike data to disk, enabling very long runs (unlimited except
for available disk space). A typical simulation of 44 hours ran in approximately
real time and produced around 2~GB of spike data. The model was implemented in
NEURON 7.2 \cite{carnevale06} for Linux, and is available on ModelDB
at the following URL: (\url{https://senselab.med.yale.edu/modeldb/enterCode.asp?model=147141}).

\textbf{Data analysis}
Simulation spike-trains were organized into multiunit activity (MUA) vectors, defined for
a cell population as the number of spikes in the population over
a time interval (bin). Bin sizes were set to 5~ms (200~Hz sampling rate).
Analyses were performed using mean-subtracted MUA vectors, with spectra calculated by the multitaper spectral
power estimation method, as implemented in the Python wrapper of the FORTRAN MTSpec library \cite{prieto2009fortran}.

\section{Simulation Results}

% %%% WL -- I usually put something here indicating the amount of work that was
% done -- eg '100 simulations were explored ...'

\subsection{Scaling Prolonged Activity During Deletion}
\begin{figure}[t]
  \centering
    %\subfloat[E activity without scaling (green/lower), with scaling (blue/upper)]{
      \includegraphics[trim = 10mm 5mm 15mm 12mm, clip,
      width=0.48\textwidth]{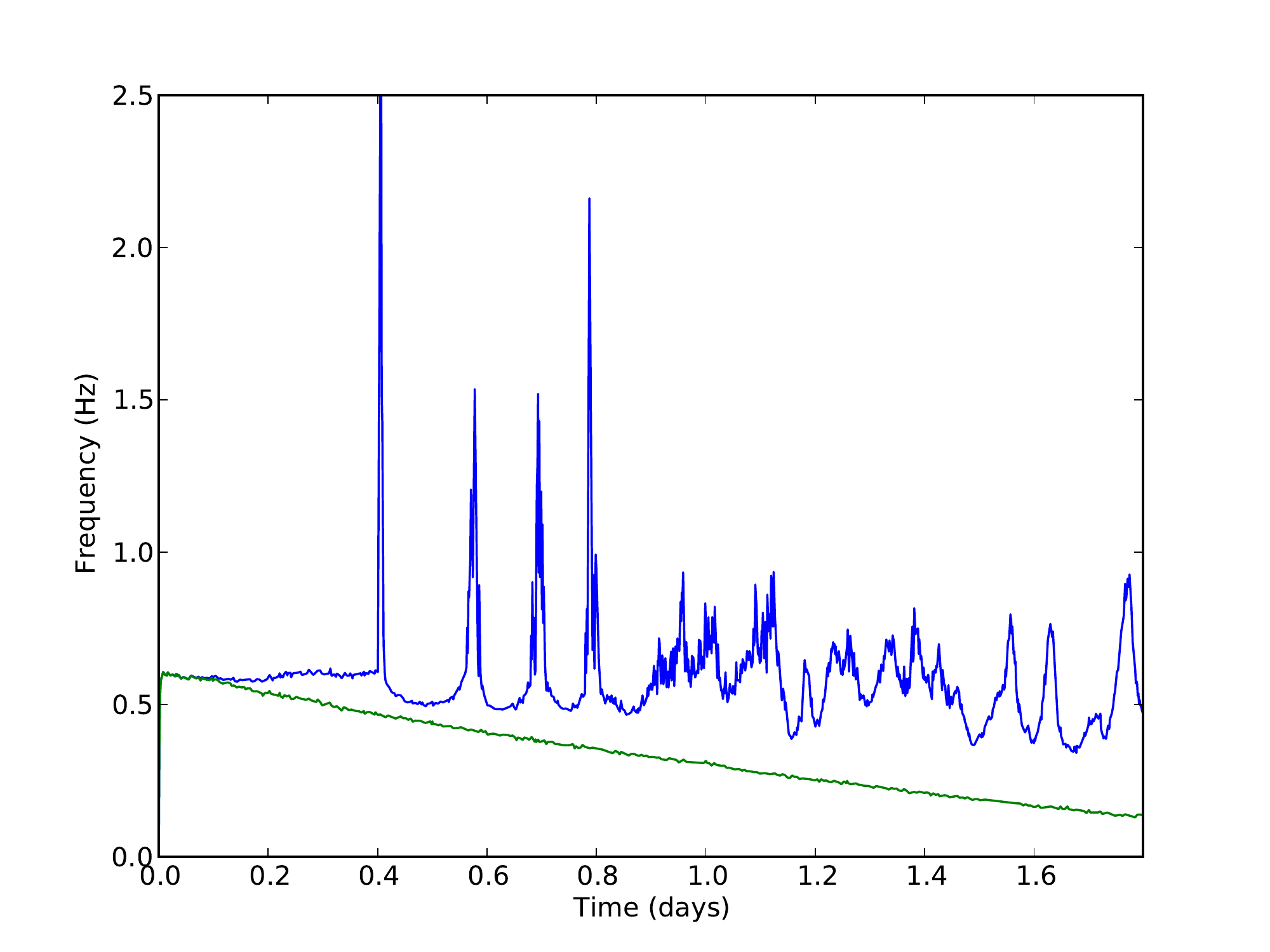}
      %\label{fig:onlydeletionact}
    %} \subfloat[Scale factors of E cells (log-scale)]{
    %  \includegraphics[trim = 8mm 5mm 15mm 12mm, clip,
    %  width=0.48\textwidth]{scalingwithdeletionscl}
    %  \label{fig:onlydeletionscl}
    %}
    \caption{E activity during pruning with (blue / upper) and without (green /
    lower) compensatory synaptic scaling. Run time 160,000~s ($\approx$ 44~h).}
  \label{fig:scalingwithdeletion}
\end{figure}

In an initial experiment, we demonstrated the usefulness of synaptic scaling by
altering network dynamics through gradual removal (pruning) of cells
(Fig.~\ref{fig:scalingwithdeletion}). Every 1600~s, three I or E neurons were
selected at random and removed from the network by setting all their synaptic
weights to zero. The global external input weights were scaled down
proportionally to the amount of deletion, at a quarter of the deletion rate, to
prevent the external inputs from swamping internal activation and artificially
raising activity. By the end of the simulation, approximately two thirds of the
cells in the network had been deleted.
%%%%%%% SN - is this mentioned in the
%%%%% methods too? TODO MR - no, as it's only applicable for this single experiment
%%%%% rather than being a general method. Do you think it should be moved to
%%%%% Methods?
%%%% WL -- if you want this paper to be about AD then have to start with some AD questions/hypos in the intro
%%%% given the readership of this journal, interest in AD somewhat limited? -- perhaps a paper on AD is next?
%%%%  (In neurodegenerative disorders such as Alzheimer's disease, it is reasonable to
%%%%  assume that activation from other cortical columns, represented in the model by
%%%%  the external inputs, would be reduced as the disease progresses). %%% SN - can
%%%%% probably use a reference for that parenthetical... TODO MR - Hmm.. it seems
%%%%% pretty obvious to me (neurons around the brain are dying -> global activation
%%%%% is reduced) but I'll see if I can find anything :)

In the absence of scaling, average firing across E cells declined steadily as
cell deletion progressed (Fig.~\ref{fig:scalingwithdeletion} green / lower
line). With scaling present, firing activity was maintained
(Fig.~\ref{fig:scalingwithdeletion} blue / upper line), with brief activity
peaks caused when the inherent delay in the activity sensor led to
over-compensation. These activity peaks do not correspond to deletion times, but
rather to emergent instabilities in the resulting damaged network. Indeed, the
network remains stable for nearly half a day following the onset of deletion
after 800~s. The over-compensation can be adjusted to some degree, although not
eliminated, by altering the scaling parameters $\beta$ and $\gamma$ (not shown).

%%%%%%% SN - in part (a) why does the E cell activity decay gradually and then
%%%%%%% suddenly rise at the end?
%%%%% MR - Looking at the rasters, it seems to be because the I cells (of which
%%%%% there were fewer to start with anyway) are now so few that they can't prevent
%%%%% the remaining E cells from increasing their activity. I'm not sure if this
%%%%% needs reporting on, as the whole hypothesis is that the green line represents
%%%%% what *doesn't* happen in the brain, as the brain utilises scaling to maintain
%%%%% activity.
%%%% WL -- the easiest way to handle this is to simply cut off the figure at 1.5 to avoid this anomaly -- 
%%%%       otherwise it is natural for the reviewers to raise questions about this
%%%%       I'm not sure what you are trying to show in panel b but i would suggest removing it
%%%%       if there is something to said about this it can be said verbally since no complex behavior here

\subsection{Synaptic Scaling Did Not Disrupt Network Behavior}
\label{sec:onlyscaling}
\begin{figure}[t]
  \centering
    \subfloat[E activity during scaling]{
      \includegraphics[trim = 10mm 5mm 15mm 12mm,
      clip, width=0.48\textwidth]{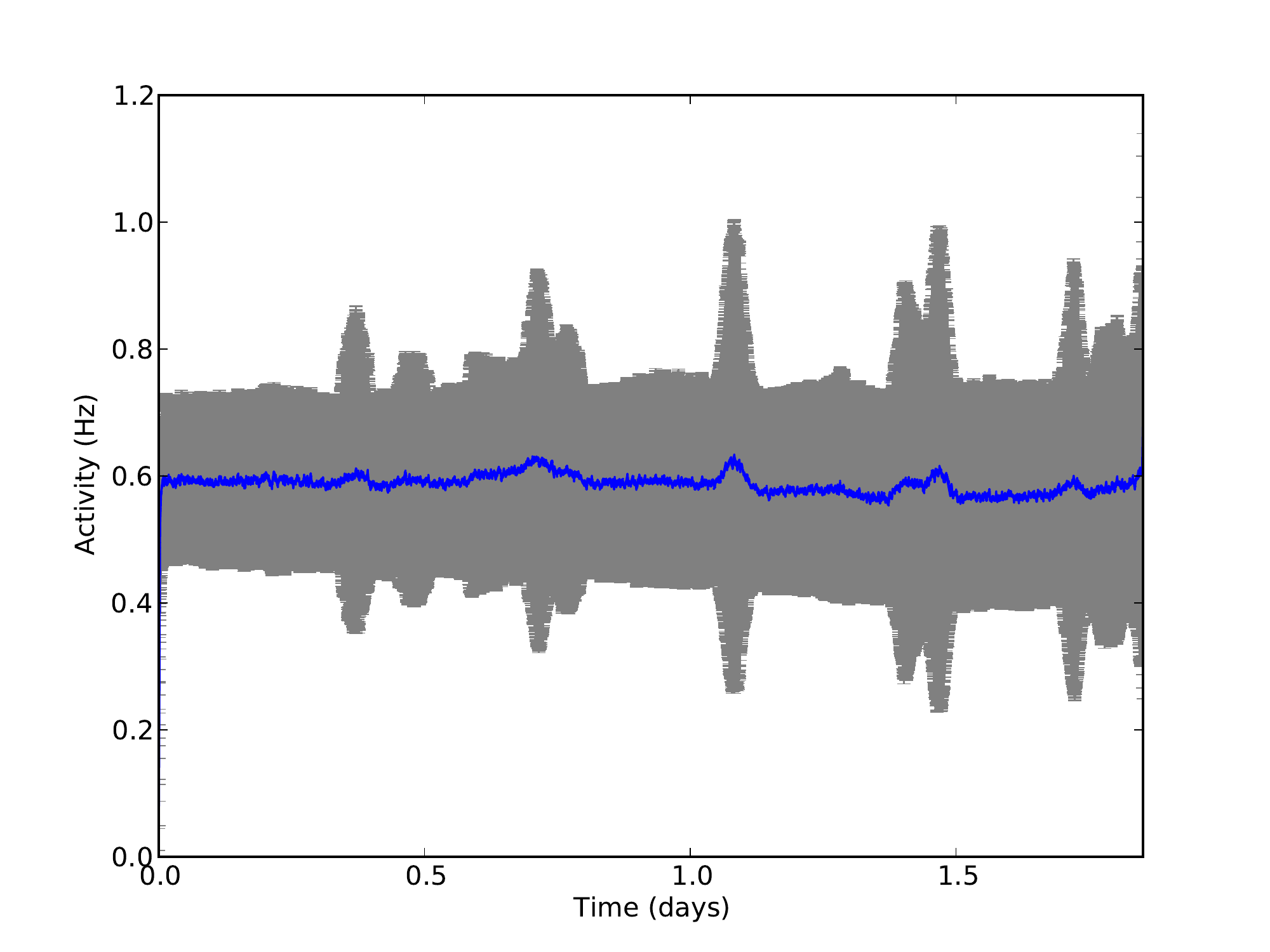}
      \label{fig:onlyscalingact}
    } \subfloat[Scale factors of E cells]{
      \includegraphics[trim = 10mm 5mm 15mm 12mm,
      clip, width=0.48\textwidth]{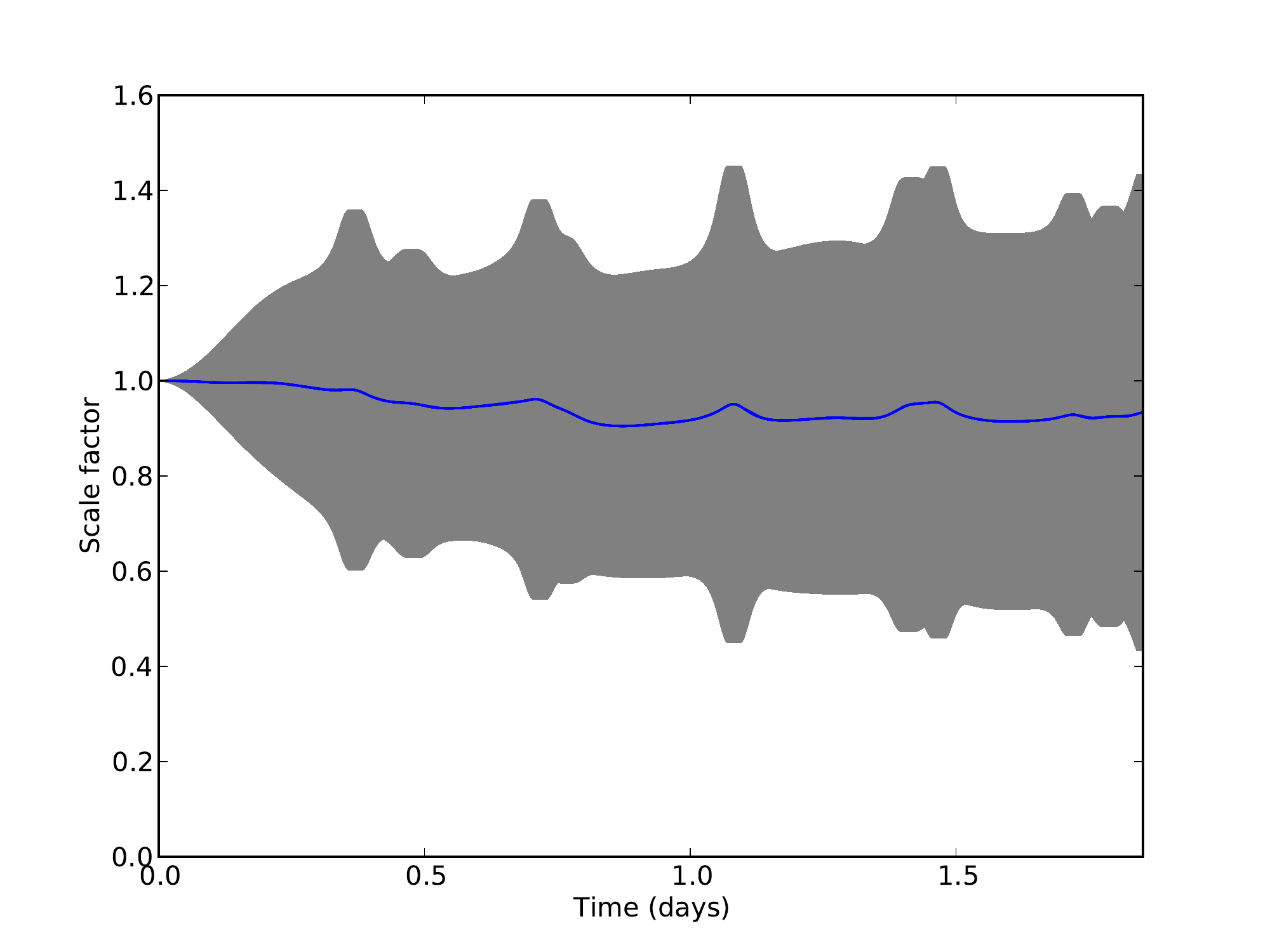}
      \label{fig:onlyscalingscl}
    }
    \caption{Scaling does not destabilize the network (mean: blue, std: grey).}
  \label{fig:onlyscaling}
\end{figure}

The model was run for 160,000~s ($\approx$~44 h) to examine the effects of
scaling over time on network dynamics. With scaling, activity of the E cells
remained steady (Fig. \ref{fig:onlyscalingact}), and scale factors remained
centered around 1 (Fig. \ref{fig:onlyscalingscl}). Scaling appeared to preserve
stability of the network during these extremely long runs.

\subsection{Unrestrained STDP Led To Hyper-Potentiation}
\label{sec:onlytraining}

\begin{figure}[t]
  \centering
    \subfloat[E activity during training]{
      \includegraphics[trim = 10mm 5mm 15mm 12mm,
      clip, width=0.48\textwidth]{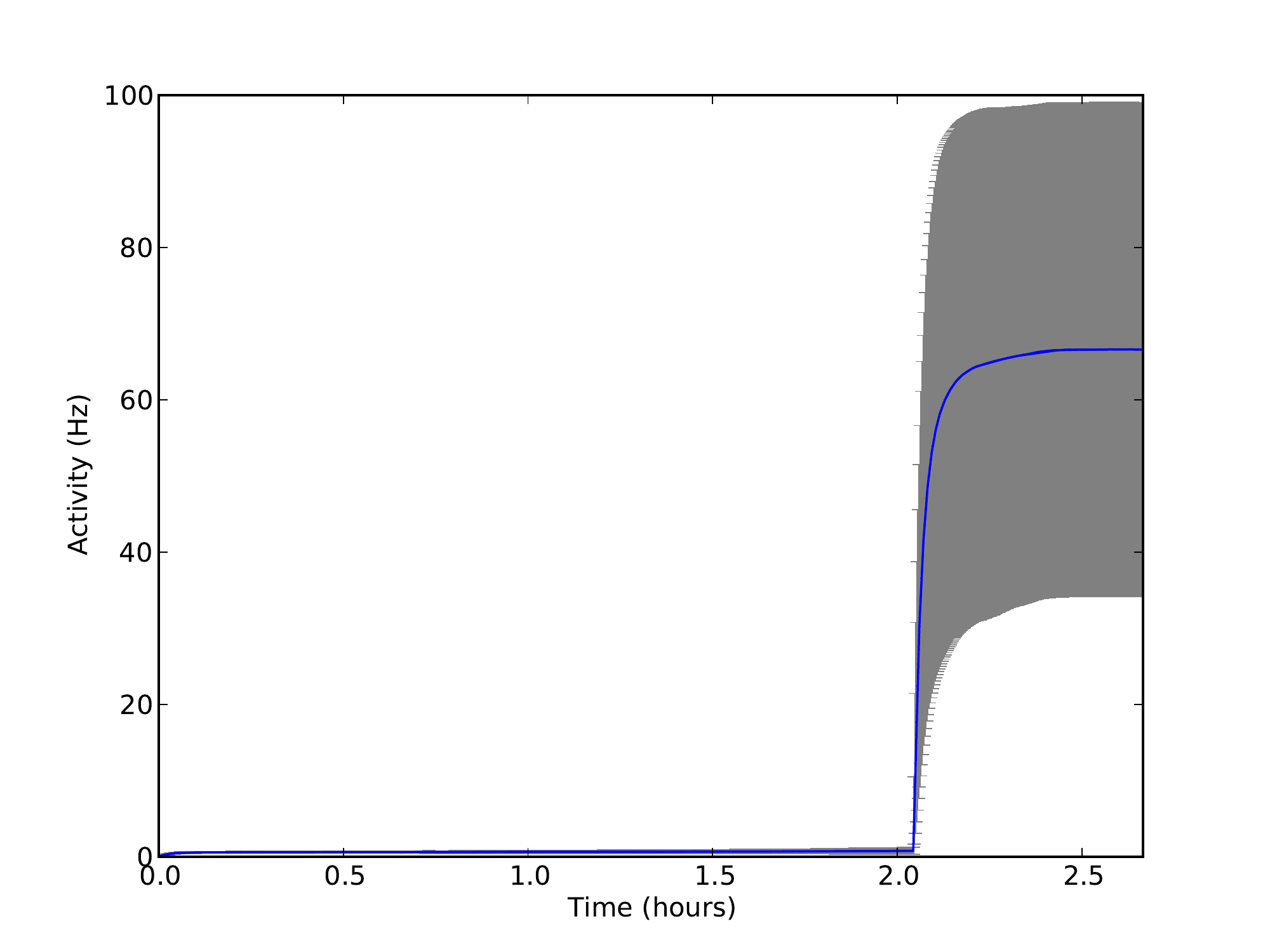}
      \label{fig:onlytrainingact}
    } \subfloat[Raster plot of 500~ms showing high-frequency network activity
    after 2~h]{
      \includegraphics[trim = 13mm 8mm 18mm 15mm,
      clip, width=0.48\textwidth]{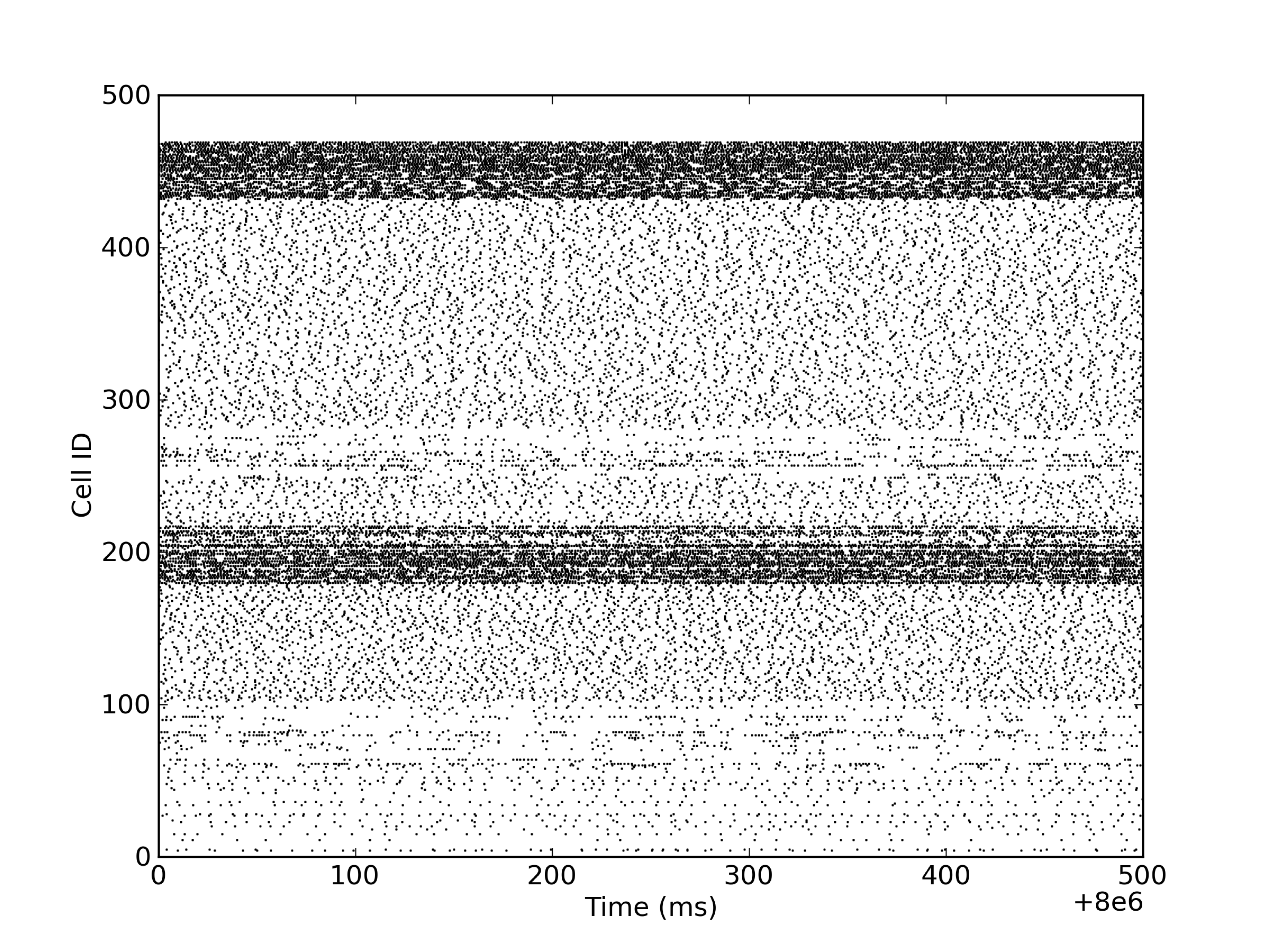}
      \label{fig:onlytrainingrast}
    }
    \caption{Training with E$\to$E STDP pushes network to high frequency activity.}
  \label{fig:onlytraining}
\end{figure}

We trained the network by applying a signal consisting of low-weight single
spikes at 8~Hz to E4 cells for 8000~s ($\approx$~2.2 h) in the absence of
synaptic homeostasis (Fig.~\ref{fig:onlytraining}). STDP was turned off during
the final 800~s in order to test recall. We found that any training signal
frequency eventually pushed the network into a state of excessive firing. This
occurred even when E$\to$I STDP balancing was added (not shown).

%%%% WL - would suggest consolidating Fig fig:powerstart and fig:powertrnscl2 into a single fig [[MAIN CONCERN]]
%%%% However, the big problem with these figs is that it's not obvious what you are showing!
%%%% The stim is 8Hz -- it is hard to see 8 hz on figs that range from 0-100 and are so unfiltered
%%%% I would suggest filtering and/or going from 0-20 for all of them and then showing 7c separately or in inset 
%%%% (play around with it to see how to see)
%%%% Also please place all of the panels at the same scale 0-0.08 so that one can compare them fairly by eye
%%%% (7c will have to  be at different scale so another reason why it might be best as an inset)
%%%% The control (Fig6) looks like it already has 8Hz -- hard to tell
%%%% The 8Hz is pretty obvious in 7a,7b
%%%% The denouement of the whole paper is ... [drumroll] fig 7d but don't see much here??
%%%% I can see that it didn't blow up but i can't see that there is much if any 8Hz
%%%% so i guess the point is that the general low-freq activity is preserved but would be nice to demonstrate that
%%%% something has been learned -- ie is Fig 7d really much different than fig6? -- looks like maybe it is since 
%%%% plotted on different scales but would be good to graph on same scale and same fig (different panels) so that can compare
%%%% more readily sidebyside and really see what has happened

\subsection{Synaptic Scaling Prevented Overactivation}
\label{sec:trainingandscaling}
\begin{figure}[t]
  \centering
    \subfloat[E activity during training with scaling]{
      \includegraphics[trim = 10mm 5mm 15mm 12mm,
      clip, width=0.48\textwidth]{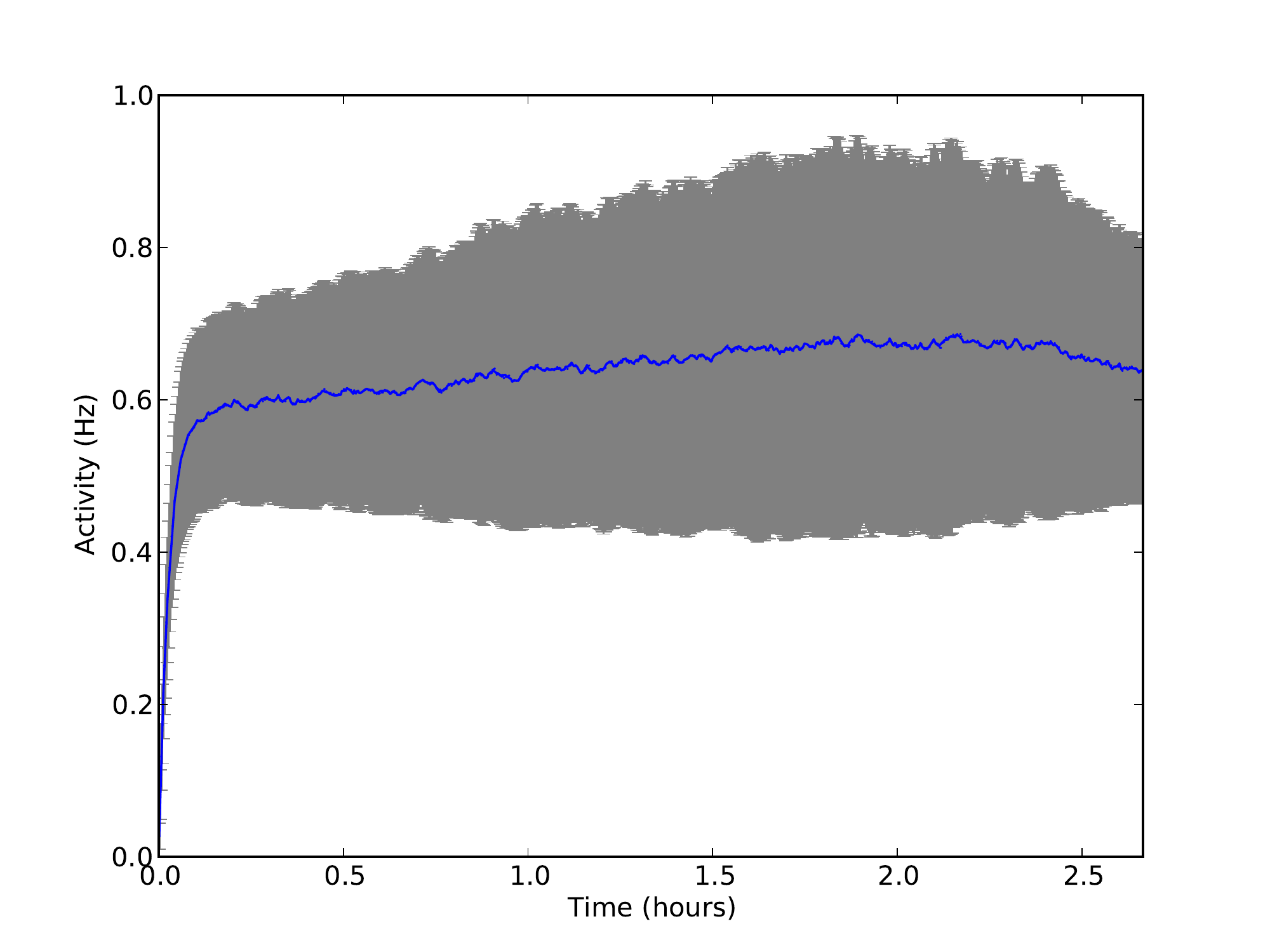}
      \label{fig:trainscaleact}
    } \subfloat[Scale factors of E cells]{
      \includegraphics[trim = 10mm 5mm 15mm 12mm,
      clip, width=0.48\textwidth]{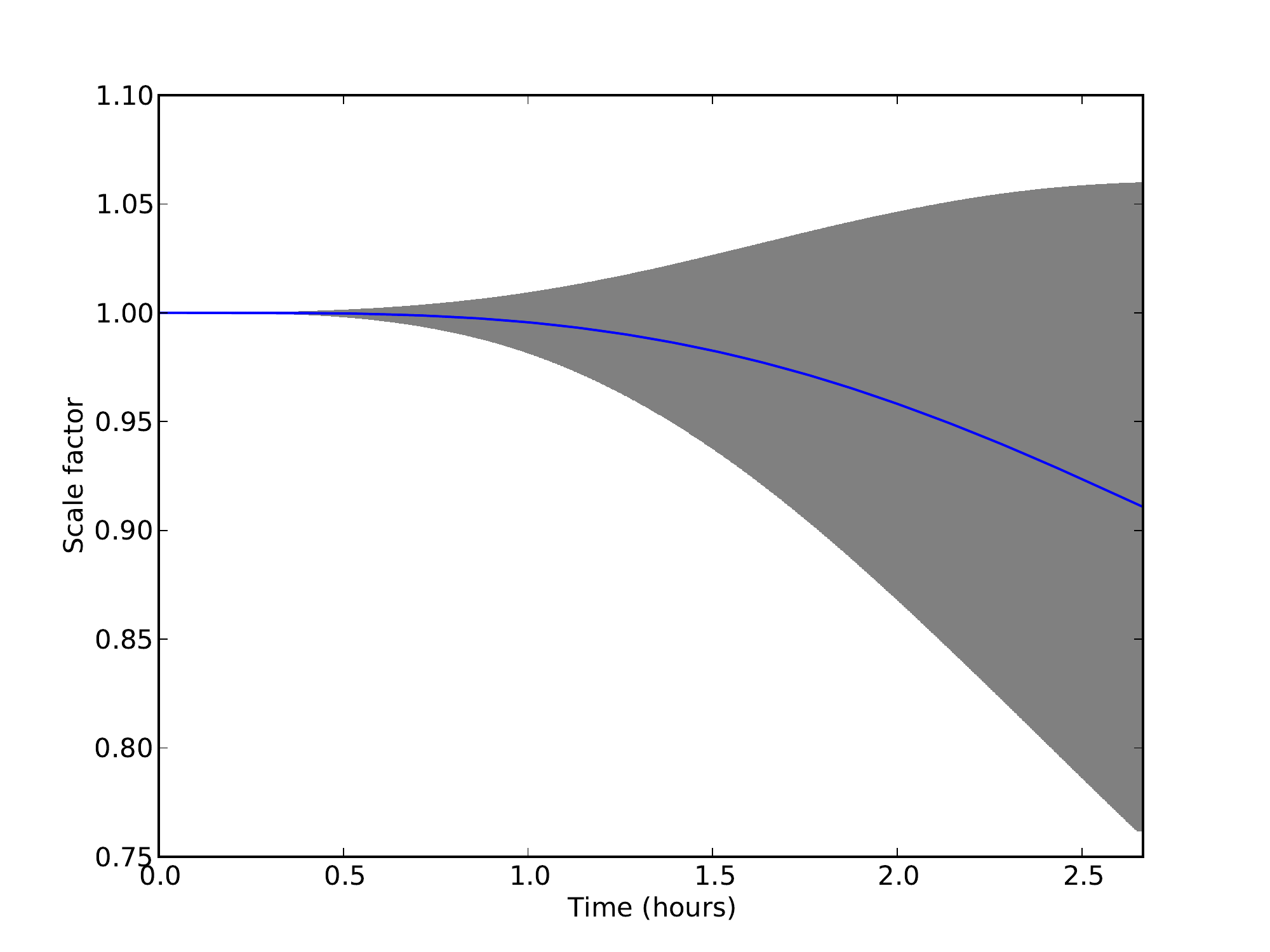}
      \label{fig:trainscalescl}
    }
    \caption{Synaptic scaling maintains E activity profile during STDP.}
  \label{fig:trainscale}
\end{figure}

We then assessed the model with STDP, training and synaptic scaling (Fig.
\ref{fig:trainscale}). Local E cell homeostatic scaling balanced the
potentiation caused by STDP, gradually scaling down all E cells, and
preventing pathological over-activation.

\subsection{Synaptic Scaling Preserved Learning}
\label{sec:preservelearning}
\begin{figure}[t]
  \centering
  \includegraphics[trim = 10mm 5mm 15mm 12mm, clip,
  width=0.5\textwidth]{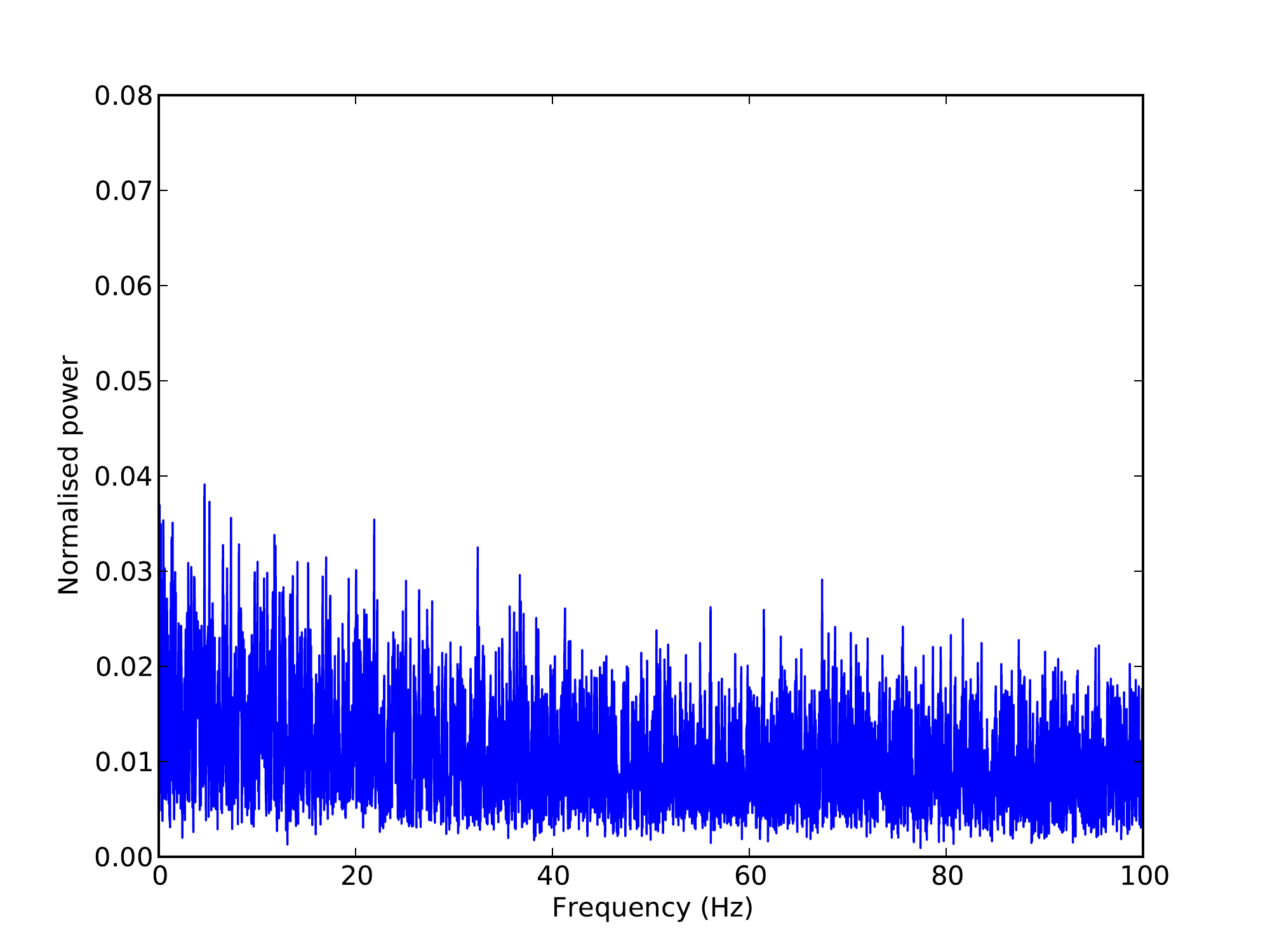}
  \caption{Baseline power spectrum of E cells (frequency (Hz) vs
  normalized power).}
  \label{fig:powerstart}
\end{figure}

Synaptic scaling served to maintain cell firing near the target rate, here the
baseline rate. However, it was possible that the scaling-down of activity would
simply reverse the potentiation caused by STDP, resulting in a loss of learned
information. In order to determine whether scaling allowed the learning of oscillations to
persist, the power spectra of the E cells were obtained at various points during
the learning process using the multitaper spectral analysis method, with spikes
sorted into 5~ms bins (Fig.~\ref{fig:powerstart}) \cite{prieto2009fortran}.
These plots show unsmoothed normalised power of the E cells within the network
at each of a range of frequencies from 0-100~Hz.

\begin{figure}[t]
  \centering
  \subfloat[With STDP only, during training]{
    \includegraphics[trim = 10mm 5mm 15mm 12mm,
    clip, width=0.48\textwidth]{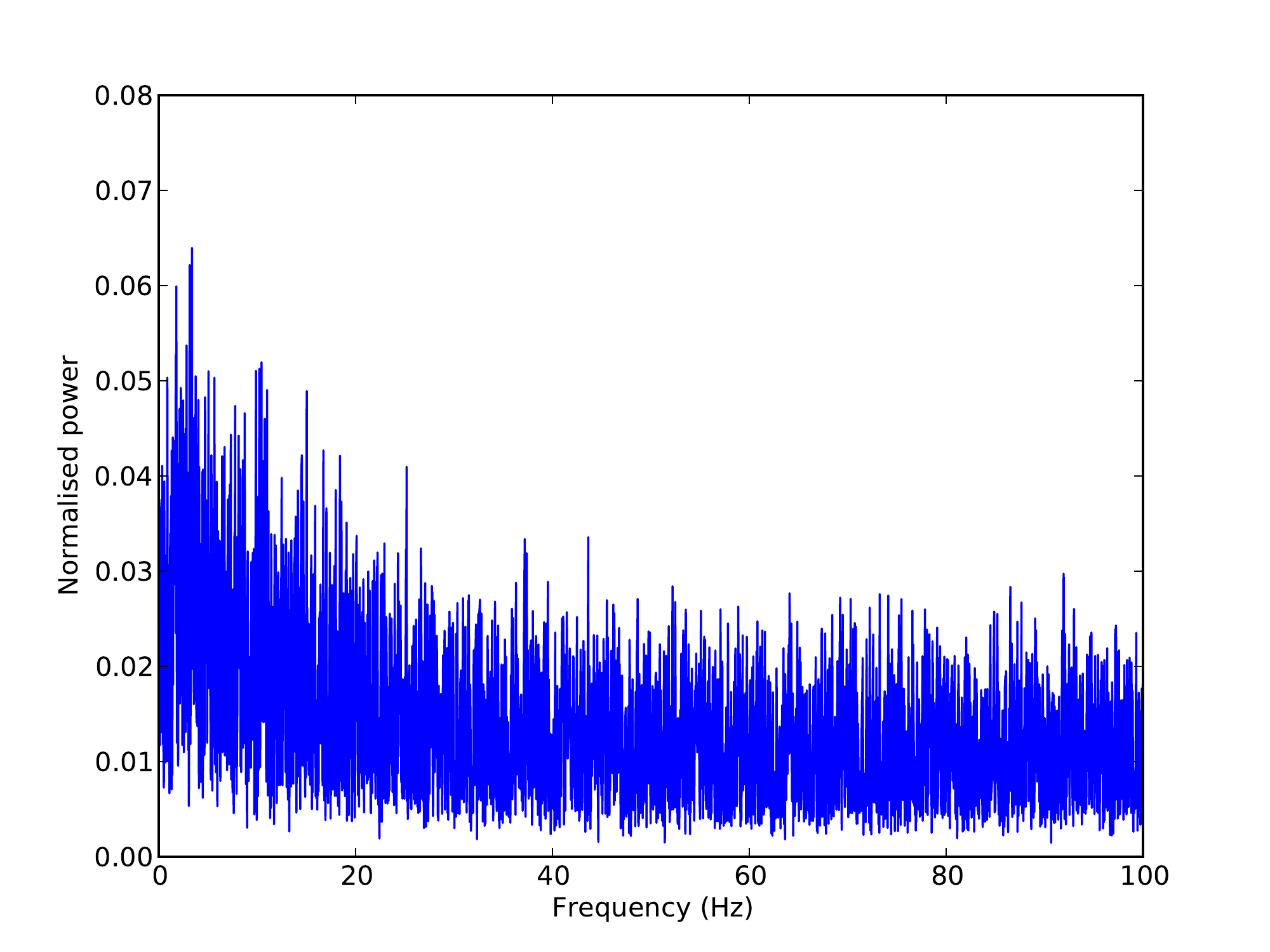}
    \label{fig:powertrn1}
  } \subfloat[With STDP \& scaling, during training]{
    \includegraphics[trim = 10mm 5mm 15mm 12mm,
    clip, width=0.48\textwidth]{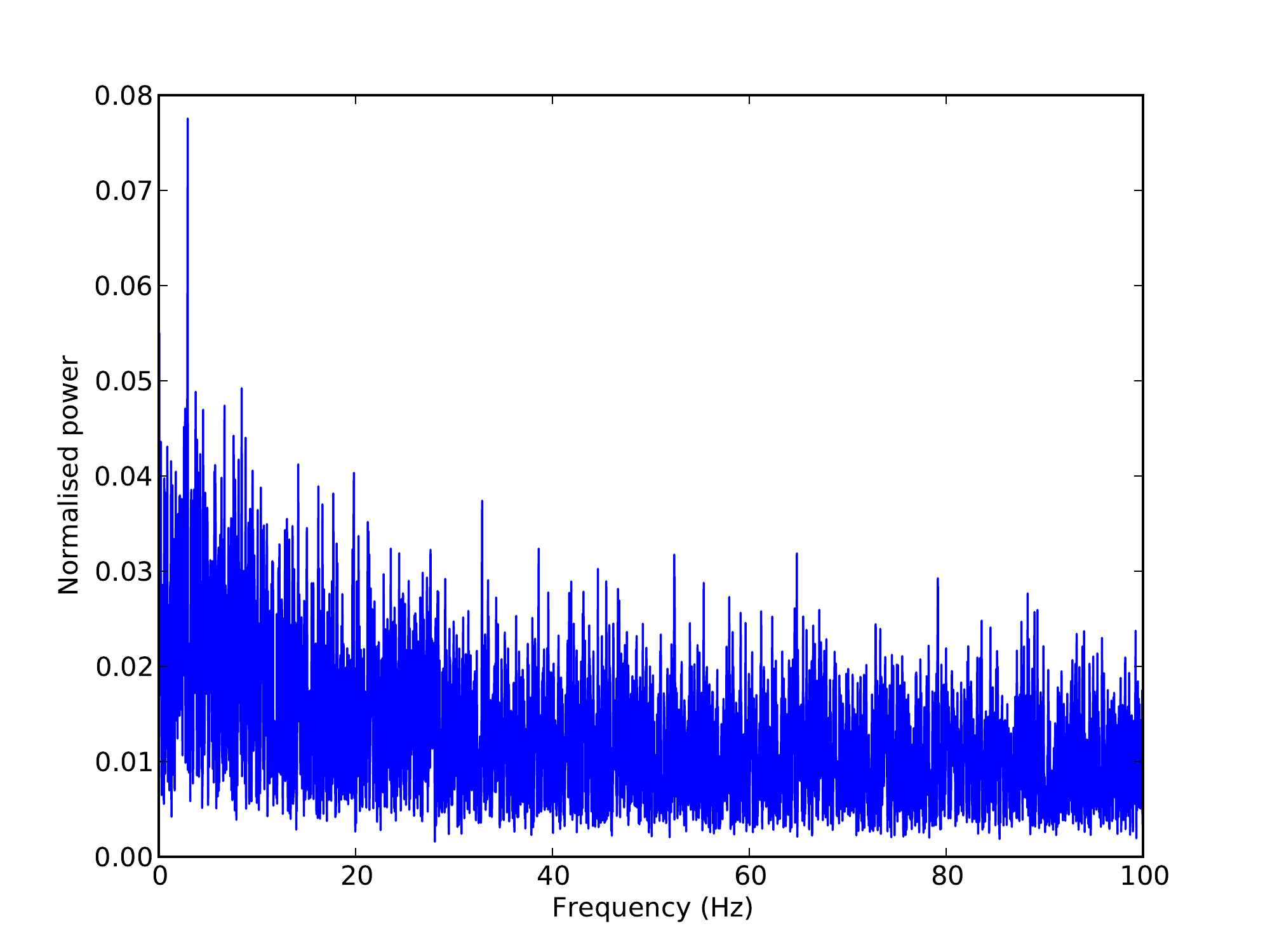}
    \label{fig:powertrnscl1}
  }
  \\
  \subfloat[With STDP only, after training]{
    \includegraphics[trim = 10mm 5mm 15mm 12mm,
    clip, width=0.48\textwidth]{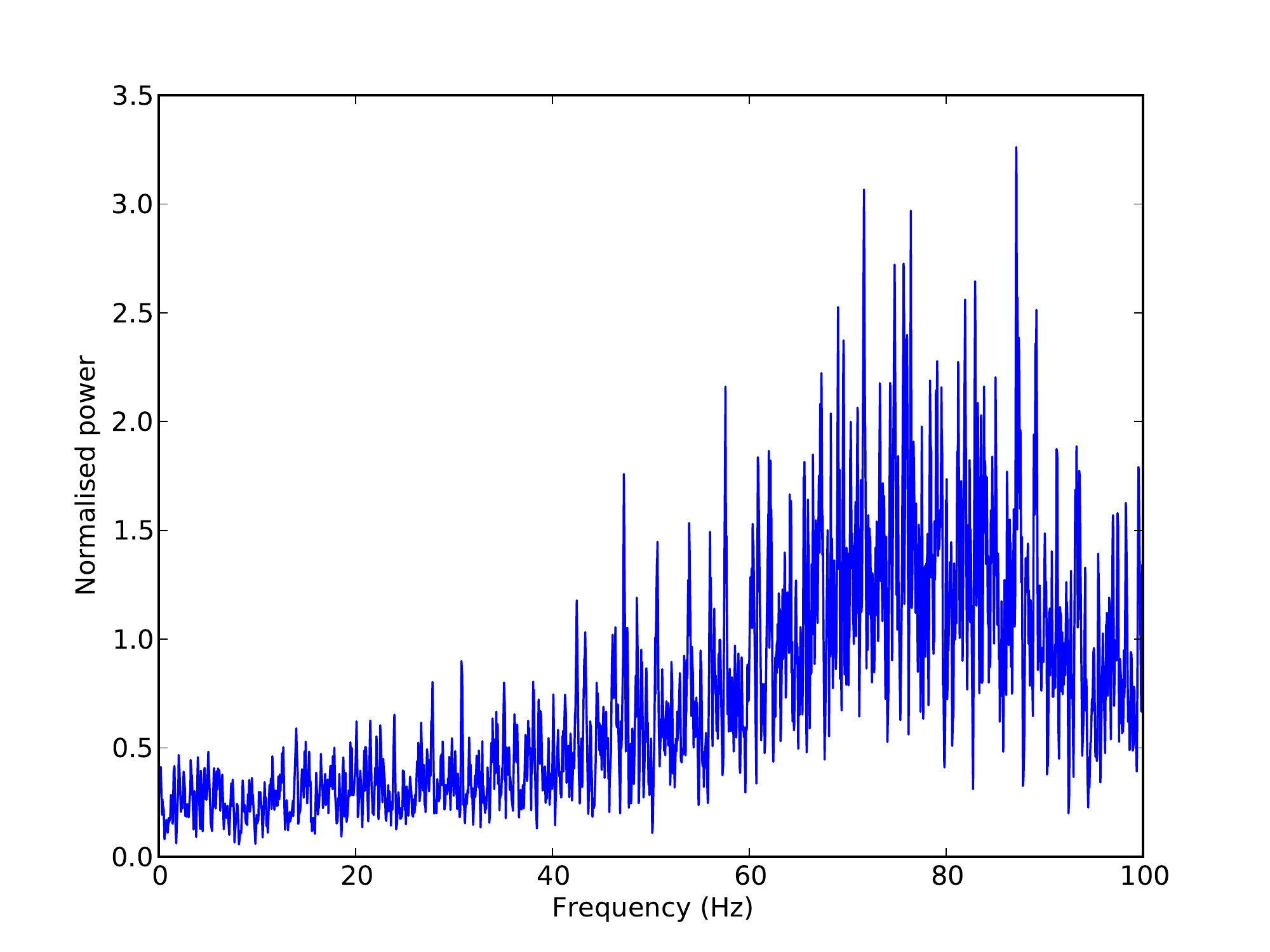}
    \label{fig:powertrn2}
  }
  \subfloat[With STDP \& scaling, after training]{
    \includegraphics[trim = 10mm 5mm 15mm 12mm,
    clip, width=0.48\textwidth]{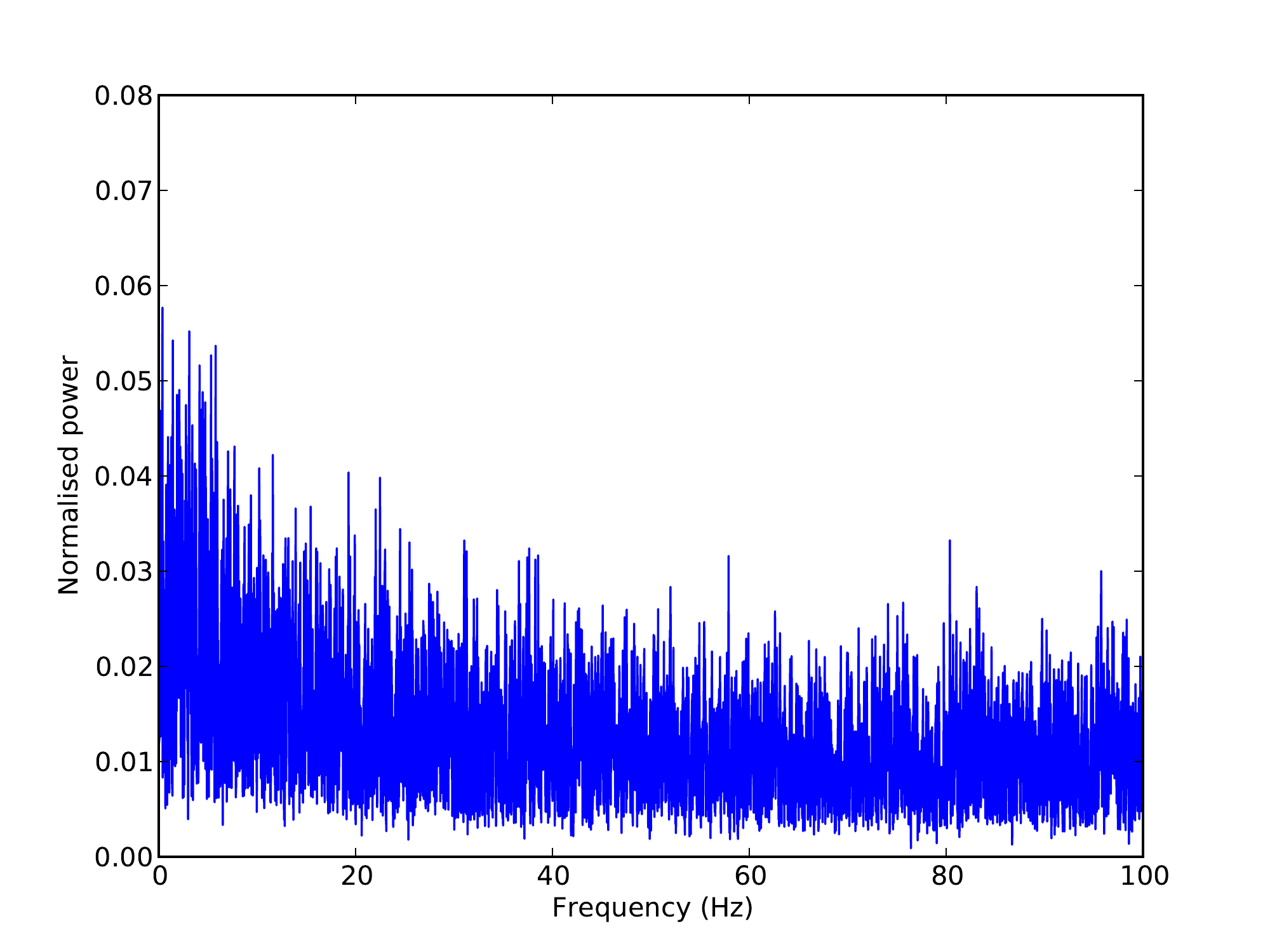}
    \label{fig:powertrnscl2}
  }
  \caption{Power spectra during (top)/after (bottom) STDP, with
  (R)/without (L) scaling.}
  \label{fig:powerspectra}
\end{figure}

STDP was applied at E$\to$E synapses for 8000~s ($\approx$ 2.2~h) with an 8~Hz
sensory signal (Fig.~\ref{fig:powertrnscl2}). In one simulation, synaptic
scaling was also switched on for E cells. Power spectra were obtained for the
period from 5600-6400~s, shortly after the middle of training (Figs.
\ref{fig:powertrn1} and \ref{fig:powertrnscl1}), and again during the recall
period at the end of learning (Figs. \ref{fig:powertrn2} and
\ref{fig:powertrnscl2}).

In both simulations, it can be seen that STDP has caused a shift in the power
spectra, with an increase in the amplitude of oscillations at low frequencies
from 0-10~Hz and a decrease above 10~Hz (Figs. \ref{fig:powertrn1} and
\ref{fig:powertrnscl1}). This demonstrates that the network has learned from the
training signal. Shortly after 7400~s (2~h), the network without synaptic
scaling transitioned to high-frequency activity, without retention of the 8~Hz
training signal (Fig. \ref{fig:powertrn2}; note different scale). However, in
the network with synaptic scaling turned on, lower frequency activity was
maintained, with a peak near the 8~Hz that was imposed during training (Fig.
\ref{fig:powertrnscl2}). Synaptic scaling therefore prevented over-activation
and preserved learning.

\section{Discussion}
This research has introduced homeostatic synaptic scaling with
dynamically-obtained target activity rates to a realistic spiking model of
neocortex which learned oscillatory frequencies via STDP. We demonstrated that
scaling is necessary for upregulation of neural activity during decline in
input. This might have implications for neurodegenerative brain disorders, in
which cortical activation might be expected to decrease. Peaks of activity were
observed during deletion due to periodic over-compensation by the scaling
mechanism. Experimental observations demonstrating hyperactivity in cells near
beta-amyloid plaques in Alzheimer's disease, and the increased incidence of
seizures in Alzheimer's patients, suggests these activity peaks may have a
biological basis
\cite{busche2008clusters,frohlich2008pathological,trasande2007activity}.
Additionally, synaptic scaling may play a significant role in the progression of
Alzheimer's disease
\cite{small2008network,rowan2012information,rowan2011effects}, and further
understanding of this mechanism and its relationship to learning and disease
pathology may be crucial to finding better treatments.

We also showed that scaling does not negatively affect the network at baseline,
but that it is stable. We demonstrated that E$\to$E scaling is sufficient to
balance the hyper-potentiation caused by unrestrained STDP. Potentiation
strengthens the co-incident connections between neurons in a positive feedback
cycle, eventually leading to hyper-potentiation, but scaling acts to shift the
mean activation constantly back towards the target activity. At the same time,
the relative (learned) distribution between postsynaptic weights remains
unaltered by scaling, and we subsequently demonstrated this principle by showing
that learning of an 8~Hz oscillatory signal is not erased by scaling.

This model investigated training and scaling at E$\to$E synapses between E
cells. While there is some evidence of STDP in I cells \cite{lamsa2010spike}, I
cells do not appear to perform scaling, but rather: ``homeostatic regulation of
inhibition is a noncell-autonomous process that either requires changes in both
pre-and postsynaptic activity simultaneously or is triggered by global changes
in network activity'' (Turrigiano et al., 2011 \cite{turrigiano2011too}). In our
model, directly enabling synaptic scaling in I cells was found to lead to
dramatic instabilities in the network dynamics (even when operating the network
at baseline, i.e. without STDP or a sensory signal), which is consistent with
Turrigiano's observations. Rather, the network appears to be most stable when I
cells are allowed to adjust their activity passively according to the changing
output from neighboring E cells, thus requiring only one dimension for the E/I
balance rather than needing a second simultaneously active dimension for
scaling.
% %%% % SN - so passive refers to adjustment of inhibition? % MR - Yes

STDP was implemented using an incremental step of $0.1\%$ of baseline synaptic
weight, which may seem low. Increasing this step size, however, meant that short
bursts of high-frequency activity were seen during learning, as the activity
sensors could not respond quickly enough to cause sufficient compensatory
scaling (although the network did soon scale back to previous firing rates).
However, 8000~s (2~h) of sustained training may also be very long compared to
biological learning from hippocampal backprojections, which is known to include
periods of recall and consolidation between periods of learning
\cite{mcclelland1995why}. This would make an interesting avenue for future
research.

%%%%% SN - low - is there a reference supporting that?
%%%% MR - no, just my own feeling given that your STDP paper used 1%
%%%% increments. van Rossum has small increments, measured in changes in the
%%%% conductance of the neural membrane, so I'll check the maths and see exactly
%%%% how small. It could be that 0.1% is fine.

%%%% % SN - so would you expect multiplicative inc to have an effect on results?
%%%% MR - yes, as weight gains would increase exponentially rather than linearly
%%%% with a multiplicative model (i.e. new weight == old weight * 1.001). I would
%%%% imagine that scaling would be able to cope with this, but without
%%%% experimenting I can't be sure.

\subsubsection{Acknowledgements.}
Research funded by EPSRC, DARPA grant N66001-10-C-2008, NIH grant R01MH086638.
The authors would like to thank John Bullinaria (Birmingham) and William Lytton
(SUNY Downstate) for their helpful comments; Michael Hines and Ted Carnevale
(Yale) for NEURON simulator support; Tom Morse (Yale) for ModelDB support; and
the anonymous reviewers for their constructive feedback.

\bibliographystyle{splncs}
\bibliography{bibliography}

\end{document}